\documentclass{iau} 
\usepackage{graphicx}

\title[Astrochemistry in external galaxies] %% give here short title %%
{Astrochemistry in external galaxies: how to use molecules as probes of their physical conditions}

\author[Serena Viti]   %% give here short author list %%
{Serena Viti}

\affiliation{Department of Physics and Astronomy, University College London, Gower St., London, WC1E 6BT, UK \\
email: {\tt serena.viti@ucl.ac.uk}} 

\pubyear{2015}
\volume{315}  %% insert here IAU Symposium No.
\jname{From interstellar clouds to star-forming galaxies: universal processes?}
\editors{xxxxx}
\begin{document}

\maketitle

\begin{abstract}
It is now well established that chemistry in external galaxies is rich and complex. In this review I will explore whether one can 
use molecular emissions to determine their physical conditions. There are several considerations to bear in mind when using molecular emission, and in particular molecular ratios, to determine the densities, temperatures and energetics of a galaxy, which I will briefly summarise here. 
I will then present an example of a study that uses multiple chemical and radiative transfer analyses in order to tackle the too often neglected
`degeneracies' implicit in the interpretation of molecular ratios and show that only via such analyses combined with multi-species and multi-lines high spatial resolution data one can truly make molecules into powerful diagnostics of the evolution and distribution of molecular gas.  
\end{abstract}
\section{Introduction}
The first detections of molecular emission in external galaxies
were made in the 1970s. Since then it has become clear that chemistry in
external galaxies can be complex and that, although most molecular emission is probably extended, it is certainly not all co-existent. CO emission is routinely used to trace the molecular mass of a galaxy, as it traces H$_2$ at large scale (see review by \cite[Bolatto et al. (2013)]{Bol13}),
although one notes that the $X$-factor used in the Milky Way to convert CO integrated line intensities to H$_2$ column densities may give misleading results if applied to other galaxies (e.g. \cite[Bell et. al. 2006]{Bell06}).
However, it is the observations of molecules with a high critical density that allow us to
trace the 
reservoir for star formation or describe the process of star formation itself, including
determining the influence of newly formed stars on their environments.

The objective of this review is to establish how one can use 
molecular emissions of different species to explore 
the physical conditions of galaxies and their likely evolutionary status, as we routinely do for our own Galaxy. In other words can astrochemistry provide the tools to explore the physical conditions and energetics of external galaxies?

There are two main challenges for the astrochemist. The first challenge is that
even for the closest 
galaxies (e.g. M82) with the most powerful interferferometer to date (ALMA), the highest spatial resolution that one can achieve is $\sim$ 0.2 pc which is still larger than an individual star forming core. In fact, for most galaxies a much worse resolution is achieved, with entire star-forming regions being unresolved (e.g. for NGC~1068, at a distance of 14 Mpc, the highest spatial resolution achieved is 30 pc, \cite[Garcia-Burillo et al. 2014]{Garc14}).
Hence the telescope beam will usually encompass 
emissions from many types of region and the molecular emission will be an 'ensemble' of multi-phase gas where the spatial and temporal effects are diluted in the beam. This will challenge the interpretation of molecular line emission, especially for galaxies very different from our own.  
The range of physical parameters (gas densities,
UV fields, cosmic ray ionization rates, dust properties, etc.)
will differ across galaxies 
(\cite[Williams \& Viti 2013]{willvi13}). 
Hence we can not necessarily 
use information about the Milky Way as a reliable guide to the
properties of other galaxies.

The second challenge for the astrochemist is the fact that molecular spectra of nearby galaxies are now 
proving to be almost as rich as those of the Milky Way (\cite[Aladro et al. 2011]{alad11}, \cite[Aladro et al. 2013]{alad13}, \cite[Costagliola et al. 2015]{costa15}). For example, an unbiased molecular line
survey in the wavelength range 1.7--2.3 mm towards the nuclear region of the
starburst galaxy NGC~253 shows the presence of more than 20 different molecular
species (\cite[Martin et al. 2006]{martin06}), all with relatively high abundances (see Table 6.2 in \cite[Williams \& Viti 2013]{willvi13}).
The concomitant detection of, say, CO, SiO, and CH$_3$OH emissions
in a spatially unresolved galaxy does not necessarily mean that they are emitted from the same gas:
the first molecule may indicate the presence of
cold diffuse clouds, the second strong shocks, and the third dense
star-forming cores. Hence
such chemical complexity cannot be explained
by a one-component model.

In order to stand up to these challenges the interpretaation of molecular emission must take into consideration the chemical routes to the formaton and destruction of each species, as well as their sensitivity to the physical and chemical conditions of the gas.  

\section{How to interpret molecular observations in external galaxies}

Molecular observations are an ideal tool to trace a wide range of densities and energetics 
in the interstellar medium of a galaxy, largely because of the wide range of
critical densities across the different molecular species and across the transitions
of the same molecule, as well as the dependendencies of chemical reactions on the energy available to the system i.e the kinetic temperature of the gas. In fact
star formation as well as AGN activity 
are both affected by the molecular gas available to the system.
 
Many studies have shown that molecular ratios, such as HCO$^{+}$/HCN or HCN/CO
differ across different types of galaxies expecially between AGN-dominated galaxies and starburst-dominated galaxies. However the derived abundance ratios for an individual galaxy also highly differ across studies depending on the transitions observed, the available resolution, and the method used in deriving the column densities.  

The differences between molecular ratios in starburst or AGN galaxies as compared to our own Galaxy 
have been interpreted as anomalous at times and an indication either of high
ionisation rates or of high star formation rates. On the other hand, it is likely that
multiple components, each with different physical conditions, are contributing
to the emission (e.g. Giant Molecular Clouds, where star formation take place, will have average number densities of the order of 10$^3$ cm$^{-3}$ and T=50-100 K, while the star-forming clumps may
have number densities up to $\sim$10$^7$ cm$^{-3}$ and temperatures as low as 10 K), so that a knowledge of the individual components is required
to interpret such ratios. This will also explain why the use of different transitions in deriving the abundance ratios as well as different beam sizes lead to substantial differences within the same galaxy. 
An understanding of the chemistry behind each molecule and its dependencies on the density and temperature of the gas will lead to a more coordinated
approach to understanding the nature of galaxies by using molecular line
emissions.

In fact it is well know that the degeneracy
between density and temperature implies that single transitions of one molecule
should not be used to determine quantitatively the gas density or temperature, especially in
extragalactic environments. Chemical models show that almost any density
can be representative of a large range of environments; for example, a density of $\leq$ 10$^4$ cm$^{-3}$  may indicate an ensemble 
of dark/quiescent gas, potential site of future star foramtion; on the other hand these are also typical densities of PDRs. 
An average density of 10$^5$ cm$^{-3}$ may indicate that star forming gas is dominant, but it may also be a signature of shocked-compressed gas. Finally a high density of $\sim$ 10$^6$ cm$^{-3}$ may be typical of left over gas from episodes of starbursts or highly shocked gas driven by AGN (\cite[Krips et al. 2011]{kri11}; \cite[Viti et al. 2014]{viti14}). In Figure 1 two chemical models showing the fractional abundance of selected species as a function of cosmic ray ionization rate are plotted: the top one represent a steady-state calculation using a PDR gas-phase code (\cite[Bisbas et al. 2012]{bisbas12}) at a density of 10$^5$ cm$^{-3}$; the bottom figure shows the results from a time dependent gas-grain chemical model (\cite[Viti et al. 2004]{viti04}) which is run in two phases: the first phase follows the collapse from a diffuse gas to a density of 10$^5$ cm$^{-3}$. The temperature is kept constant at 10 K and atoms and molecules freeze on to the dust and are involved in surface reactions. The second phase follows the gas as its temperature is increased to 100 K due to star formation taking place, with subsequent sublimation of the icy mantles formed during the first phase. The two models are clearly very different and simulate physically different environments; and yet
for the same cosmic ray ionization rate, some of the abundances are remarkably similar: take for example the HCN/HCO$^+$ at a cosmic ray 100 times that of our own Galaxy. Clearly, this ratio alone would not be able to inform the observer whether it is tracing star forming gas, AGN or PDR dominated gas without any prior information or more molecular observations. 
\nopagebreak
\begin{figure}[tbh!]
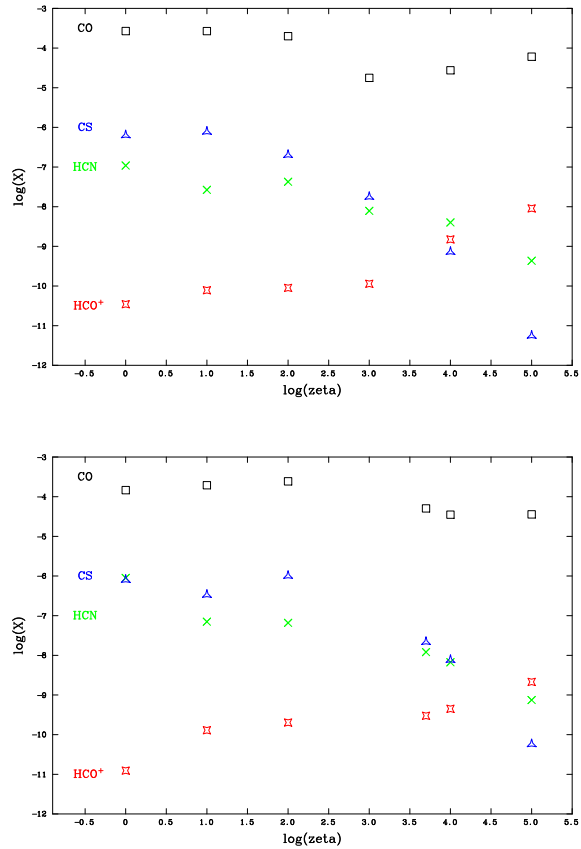

% \vspace*{-2.0 cm}
\begin{center}
 \includegraphics[angle=-90,width=3.4in]{Viti.Fig1a.eps}
\includegraphics[angle=-90,width=3.4in]{Viti.Fig1b.eps}
% \vspace*{-1.0 cm}
 \caption{Fractional abundances (with respect to the total number of hydrogen nuclei) of selected species as a function of cosmic ray ionization rate for two types of chemical models. See text.}
   \label{fig1}
\end{center}
\end{figure}

\subsection{Why are molecular ratios 'degenerate'?}
Most molecular species have more than one route of formation and destruction. For example, let us consider a collapsing core undergoing collapse from a diffuse state, e.g. 100 cm$^{-3}$, to a high density of, say, 10$^7$ cm$^{-3}$, under standard interstellar conditions and let us look at the chemistry of HCO$^+$ and HCN, the two most commonly observed extragalactic species, after CO. At very low densities the following routes of formation are important:

\begin{equation}
{\rm H} + {\rm HCN}^+ \rightarrow {\rm HCN + H}^+ 
\end{equation} 

\begin{equation}
{\rm C}^+ + {\rm H}_2{\rm O} \rightarrow {\rm HCO}^+ + {\rm H}^+ 
\end{equation}

while at intermediate densities:

\begin{equation}
{\rm N} + {\rm CH}_2 \rightarrow {\rm HCN + H} 
\end{equation}

\begin{equation}
{\rm C} + {\rm NH}_2 \rightarrow {\rm HCN + N}   
\end{equation}

and

\begin{equation}
{\rm H}_3^+ + {\rm CO} \rightarrow {\rm HCO}^+ + {\rm H}_2
\end{equation}

\begin{equation}
{\rm CH} + {\rm O} \rightarrow {\rm HCO}^+ + {\rm e}^{-}
\end{equation}

contribute $\sim$ 30\% each to the formation of HCN and HCO$^+$ respectively.
Finally, at high densities:

\begin{equation}
{\rm NH}_3 + {\rm CN} \rightarrow {\rm HCN + NH}_2   
\end{equation}

and Reaction 2.5, are the two dominant routes for HCN and HCO$^+$ respectively.  

The formation and destruction routes are not just dependent on the density of course. For constant density, variation in cosmic ray ionization rate can lead to different routes of formation and destruction. For example, CS, another routinely observed molecule in extragalactic environemts, is very sensitive to cosmic ray ionization rates. At $\zeta$ $>$ 100 $\zeta_0$ the main destruction mechaniscm for CS is: CS + O $\rightarrow$ CO + S, while at low ionization rates H$_3$O$^+$ + CS $\rightarrow$ HCS$^+$ + H$_2$O is dominant. 

From an observational point of view the variations in the chemistry are of course reflected in the observed line intensities. Figure 2 shows theoretical line intensities for CS for a very small grid of chemical and radiative transfer models.
\nopagebreak
\begin{figure}[tbh!]
% \vspace*{-2.0 cm}
\begin{center}
 \includegraphics[width=3.4in]{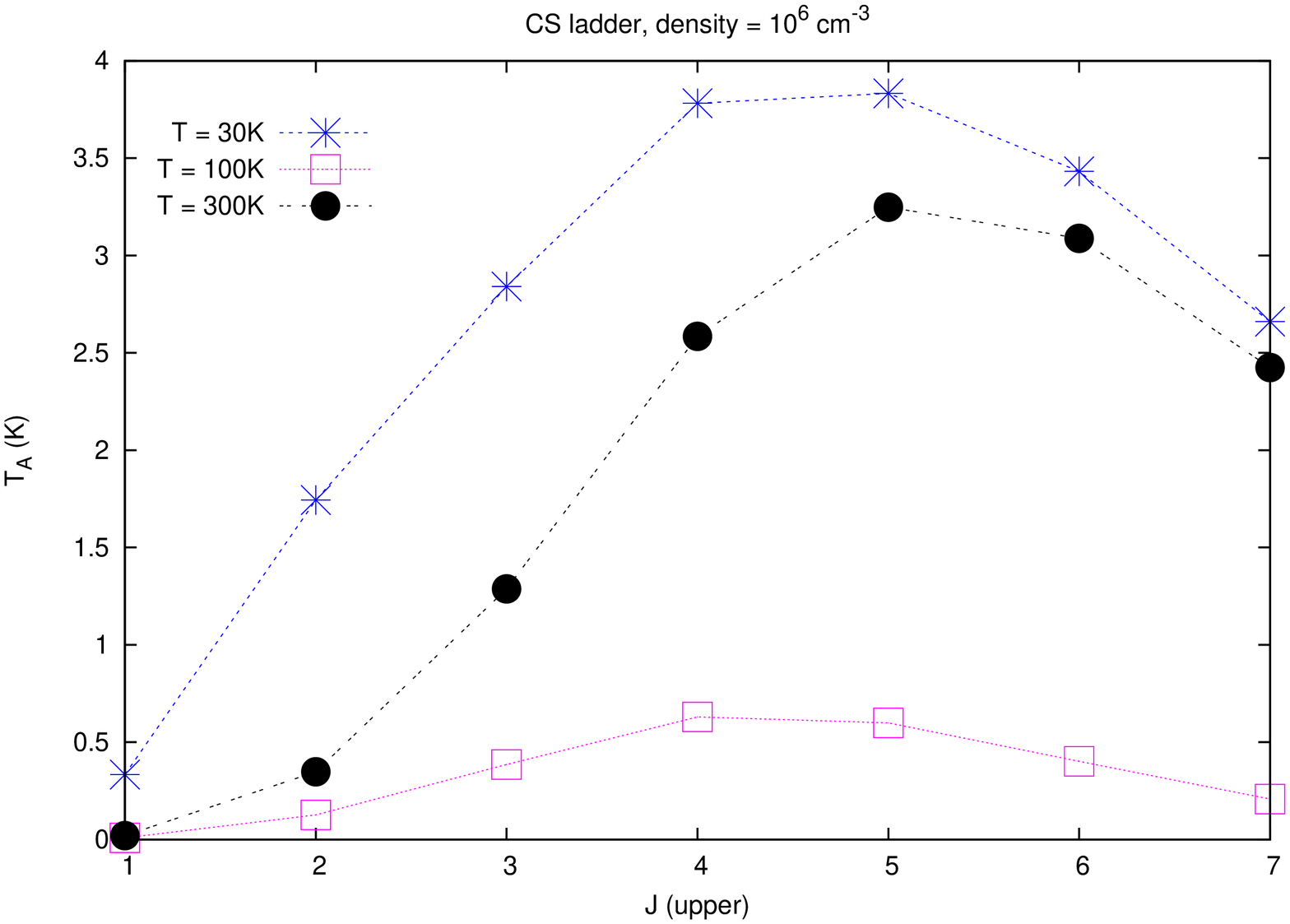}
 \includegraphics[width=3.4in]{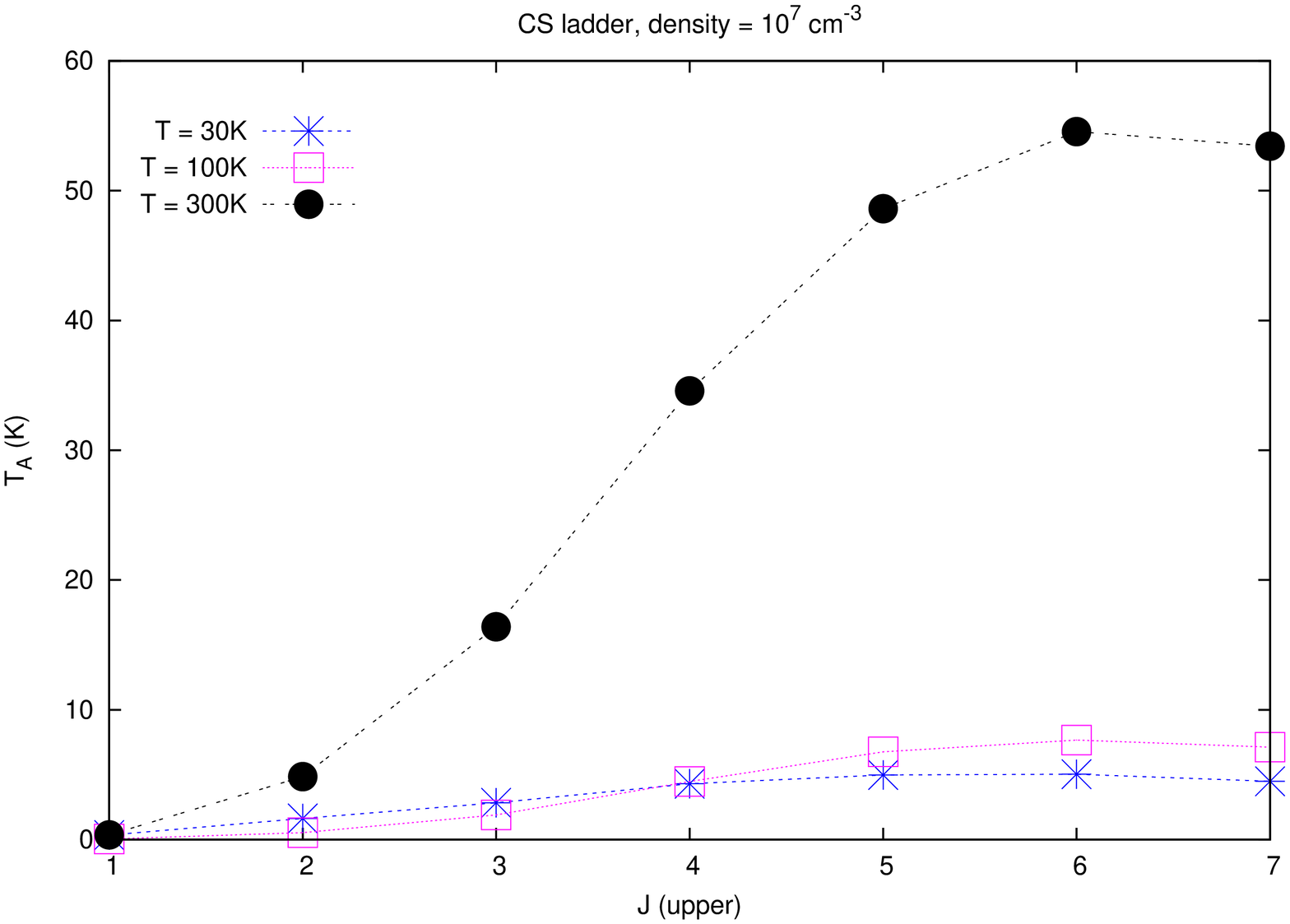}
% \vspace*{-1.0 cm}
 \caption{Theoretical CS ladders calculated by coupling chemical (UCL\_CHEM, \cite[Viti et al. 2011]{Viti11}) and line radiative transfer models (SMMOL, \cite[Kelly et al. 2015]{kelly15}).}
   \label{fig2}
\end{center}
\end{figure}
\nopagebreak

Already from such a small grid (where we
only used two gas densities and three gas temperatures, keeping constant the cosmic ray ionisation
rate and not including shocks, for example) one can see how sensitive the intensities of the CS transitions are;
it is especially interesting to see how the ladder peaks at different J depending on the gas
temperatures and densities.

When one considers that hundreds of species involved in thousands of reactions may be present in the gas, then disentangling the chemistry from the physics is not a trivial task, especially as we must not forget the role of dust and surface reactions. For example, methanol is routinely observed in extragalactic environments, and yet there is no efficient way of forming it in the gas at low temperatures (T $<$ 100 K). It is likely that most of the CH$_3$OH observed in extragalactic environments has formed on dust grains and then sublimated back in the gas phase. 

In summary, it is important to emphasize that chemistry leads to a large range of ratios, heavily dependent on small changes in the physical and chemical (initial) conditions of the gas. Hence one {\it cannot} use molecular ratios without considering of the chemistry behind them and without paying attention to individual abundances. Performing multiple, independent analyses of the observational data is therefore essential as we shall demonstrate in the next section.

\section{NGC~1068: a case study}

The unprecedented resolution of ALMA has finally allowed us to resolve structures inside nearby galaxies. NGC~1068, a composite starburst/AGN galaxy, is a Seyfert 2 galaxy, at a distance of $\sim$ 14 Mpc. Many studies have focused on this galaxy (e.g. \cite[Usero et al. 2004]{User04}, \cite[Israel et al. 2009]{Isra09}, \cite[Kamenetzky et al. 2011]{Kame11}, \cite[Hailey-Dunsheath et al. 2012]{Hai12}, \cite[Aladro et al. 2013]{alad13}, \cite[Schinnerer, et al. 2000]{Schi00}, \cite[Garcia-Burillo et al. 2010]{Garc10}, \cite[Krips et al. 2011]{Krip11}, \cite[Garcia-Burillo et al. 2014]{Garc14}, \cite[Viti et al. 2014]{Viti14}) and have shown that molecular tracers of dense gas are essential to spatially resolve the distribution, kinematics, and excitation of the circumnuclear gas of NGC~1068, as well as to study  the relationship between the $r\sim1-1.5$~kpc starburst ring and the $r\sim200$~pc circumnuclear disc (CND) located around the AGN. 
In \cite[Garcia-Burillo et al. (2014)]{Garc14} ALMA Cycle 0 observations in Bands 7 and 9 of several molecular transitions, namely CO (3-2) and (6-5), HCO$^+$ (4-3), HCN (4-3), and CS (7-6)
within the $r\sim200$ pc CND, were presented (see Figure 3). 
\begin{table}
  \begin{center}
  \caption{Table 10 from \cite[Viti et al. (2014)]{viti14}. Grid of chemical models. 
The cosmic ray ionization rate, the radiation field, the gas temperature, and the gas density are listed in columns 2--5.}
\label{tab1}
 {\scriptsize
\begin{tabular}{|c|ccccc|}
\hline
Model n$^o$ &  $\zeta$ (s$^{-1}$) & $G_0$ (Draine) & T (K) & n$_{final/preshock}$ (cm$^{-3})$ & Shock \\
\hline
1 & 1&  1 & 100 & 10$^4$  & N \\
2 & 1& 1& 100 & 10$^5$  & N \\
3 & 1& 1& 100 & 10$^6$  & N \\
4 & 10& 1& 100 & 10$^4$  & N \\
5 & 10& 1& 100 & 10$^5$  & N \\
6 & 10& 1& 100 & 10$^6$  & N \\
7 & 1& 10& 100 & 10$^4$  & N \\
8 & 1& 10& 100 & 10$^5$  & N \\
9 & 1& 10& 100 & 10$^6$  & N \\
10 & 10& 10 & 100 & 10$^4$  & N \\
11 & 1& 500 & 100 & 10$^4$  & N \\
12 & 500& 1& 100 & 10$^4$  & N \\
13 & 5000& 1& 100& 10$^4$  & N \\
14 & 10$^5$ & 1& 100 & 10$^5$  & N \\
15 & 10& 1& 100 & 2$\times$10$^6$  & N \\
16 & 1 & 1 & 100 & 5$\times$10$^6$  & N \\
17 & 1 & 1 & 100 & 2$\times$10$^6$  & N \\
18 & 1 & 1 & 200 &  10$^5$  & N \\
19 & 1 & 1 &  200 &  10$^4$  & N \\
20 & 1 & 1 & 200 & 10$^6$  & N \\
21 & 10 & 1 & 200 & 10$^6$  & N \\
22 & 1 & 10 & 200 & 10$^6$  & N \\
23 & 10 & 1 & 200 & 10$^5$  & N \\
24 & 1 & 10 & 200 & 10$^5$  & N \\
25 & 1& 1 & -- & 10$^4$  & Y \\
26 & 10& 1 & -- & 10$^5$  & Y \\
27 & 1& 1 & -- & 10$^5$  & Y \\
\hline
\end{tabular}
}
\end{center}
\end{table}

\nopagebreak
\begin{figure}[tbh!]
% \vspace*{-2.0 cm}
\begin{center}
\includegraphics[angle=0,width=3.0in]{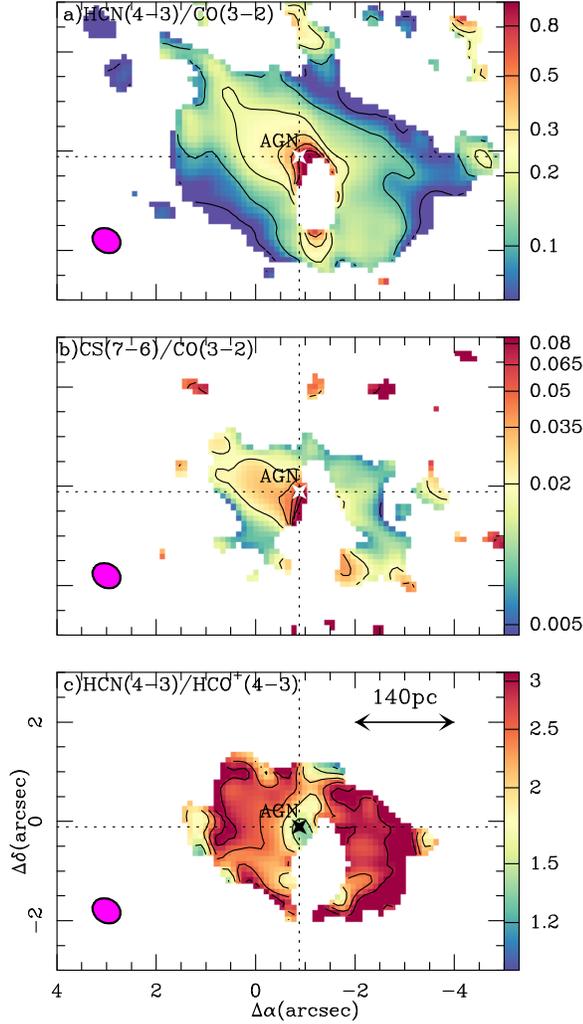}
% \vspace*{-1.0 cm} 
\caption{Figure 5 from \cite[Viti et al. (2014)]{Viti14}. HCN(4-3)/CO(3-2), CS(7-6)/CO(3-2) and HCN(4-3)/HCO$^+$(4-3) velocity-integrated intensity ratios in the CND. The filled ellipses represent the spatial resolutions used to derive the line ratio maps:  0.6$^"\times$0.5$^"$ PA = 60$^0$, for all panels.}

   \label{fig3}
\end{center}
\end{figure}
\nopagebreak
\cite[Viti et al. (2014)]{Viti14} performed a chemical analysis of the gas within the CND with the aim of quantifying the chemical differentiation and of determining the chemical origin of such differentiation: the ALMA data were used together with lower resolution IRAM Plateau de Bure Interferometer data covering the lower-J transitions of each molecule.  Five different analyses were performed: an LTE analysis, including a rotation diagram; three different analyses using RADEX (\cite[van der Tak et al. 2007]{vdt07}) differing in the number of ratios and species used; and finally a purely {\it ab initio} analysis using a time dependent gas-grain chemical mode. While each analysis led to its own detailed conclusions, all five indicated that {\it more than one gas-phase component is necessary to fit all the available molecular ratios uniquely.} Chemically this is certainly not unexpected, as discussed in the section above. Of particular interest to this review is Figure 12 in \cite[Viti et al. (2014)]{Viti14}, which we include here together with their Table 10, in Figure 4 and Table 1 respectively.
One notes first of all that the y-axis shows logarithmic values of column density ratios, and yet a very large range of ratios are displayed. None of them are anomalous in that the grid of models performed covers reasonable values of densities, cosmic ray ionozation rate, temperatures and radiation field. \cite[Viti et al. (2014)]{Viti14} found that each subregion of the CND could in fact 
be fitted by more than one set of column densities and, more importantly, that
 there is not one single model that can explain {\it all} ratios: multiple gas components even within each sub region are required. In fact, the multi-analyses performed in \cite[Viti et al. (2014)]{Viti14} seem to indicate that there has to be a pronounced chemical differentiation across the CND and that each subregion could be characterized by a 3-phase component ISM, with two gas phases at different densities but both subjected to a high ionization rate, and one gas component comprising of shocked gas where most likely the CS arises from.
\nopagebreak
\section{Concluding remarks}
Recent years have witnessed a plethora of extragalactic molecular
surveys which brought detailed information about the amount and
distribution of the molecular gas in different types of galaxies.
High resolution multi-species as well as multi-line molecular observations coupled with multiple independent analyses take us a step closer towards disentangling the gas components and energetics of nearby galaxies.
\begin{figure}[tbh!]
\begin{center}
\includegraphics[angle=-90,width=4.0in]{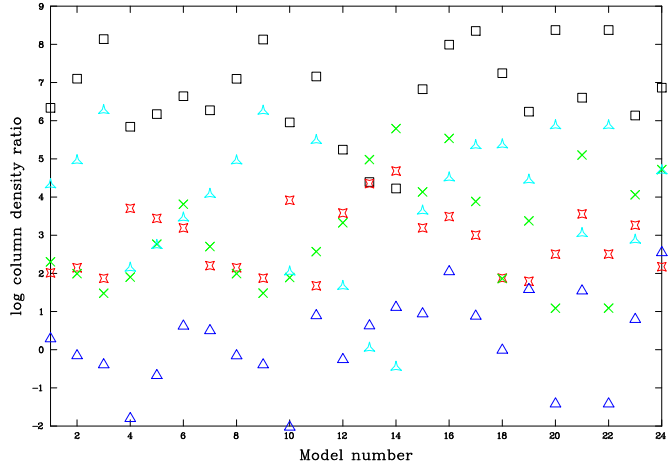}
\caption{Theoretical fractional abundances ratios for the grid of UCL\_CHEM models at high A$_V$ (10 mags). The black squares are CO/HCO$^+$ ratios; red stars = CO/HCN; green crosses = CO/CS; cyan triangles = HCN/CS; blue triangles = HCN/HCO$^+$. The numbering in the x-axis refers to the Model number (see Table 1).}
\end{center}
\end{figure}
\nopagebreak

\end{document}